# One-dimensional magnetism of one-dimensional metallic chains in bulk $MnB_4$.

S. Khmelevskyi[1*], J. Redinger[1], A.B. Shick[2], and P. Mohn[1].

[1] *Institute of Applied Physics, CMS, Vienna University of Technology, Gusshausstrasse 25, Vienna*
[2] *Institute of Physics ASCR, Condensed Matter Theory, Na Slovance 3, Prague, Czech Republic*

We have investigated from first-principles an electronic structure and magnetism in $MnB_4$ compound with experimentally observed orthorhombic C12/m1 structure. It is found that Mn tetra-borides ($MnB_4$) is found to have metallic ground state with well defined local Mn magnetic moments. This conclusion was drawn from calculation within full potential Linear Augmented Plane Wave method and Korringa-Kohn-Rostocker method using Disordered Local Moment Approximation. We have shown using Lichtenstein Green function method that magnetic exchange interactions between Mn moments are strongly ferromagnetic along 1D-chains of Mn atoms and they are practically vanishing between the chains. The metallic state appears to exhibit a strongly one-dimensional character since the single metallic band show dispersion only in one reciprocal lattice dimension. Thus it appears that MnB4 may be a perfect one dimensional one band Hubbard model system. Although LSDA predicts the magnetism in this system it may well be superconductor at low temperatures due to the correlation effects beyond mean filed like approach.

*Corresponding author: Tel:+43-158801-15838, Fax: ++43-158801-15898,
e-mail: sk@cms.tuwien.ac.at

# 1. Introduction.

The tetra-boride $MnB_4$ with body centered monoclinic structure is highest known boride of Mn [1]. The lower Mn borides are magnetically ordered [1], most known of them is high temperature antiferromagnet $MnB_2$ (see Ref.[2] and references therein). However, the magnetism in $MnB_4$ has been never reported in literature. Crystal structure of this material has been investigated in earlier 70ies [3] and then this material has attracted some attention in theoretical chemistry since it has unique and very unusual tetragonal Boron (carbon) network [4]. Monoclinic unit cell of this material has very short c-axis (2.94 A) compare to the a- (5.55 A) and b- (5.44 A) axis [3]. Along this short axis there are Mn chains, which are very well separated from each other by distance and boron network as illustrated in the figure 1.

[figure 1 here]

Here we undertake a first-principles investigation of electronic structure and magnetism in this system including calculations of the exchange constants. We have found a ferromagnetic ground state and quasi one dimensional character of the magnetism and metallicity in this system. In contrast to earlier known 1D magnets, including recently discovered metallic antiferromagnetic $NaV_2O_4$ [5], the compound $MnB_4$ is ferromagnetic and its magnetism has strongly itinerant character.

# 2. Calculational Details.

The electronic structure of $MnB_4$ compound has been calculated using the relativistic version of the full potential Linear Augmented Plane Wave (FP-LAPW) method [6] and in addition by the scalar relativistic Korringa-Kohn-Rostokker Green Function based method within the Atomic Sphere Approximation (KKR-ASA) [7]. In both sets of calculations the exchange and correlation effects are treated within the framework of Local Spin Density Approximation (LSDA) using the parameterisation by von Barth and Hedin [8]. The calculations are done for the experimental lattice structure [3].

In the case of KKR-ASA calculations we used a spherical harmonic expansion up to $l_{max}=2$ (*spd*- basis set). Multipole moment contributions to the non-spherical part of the electrostatic contribution to the one electron potential inside the ASA spheres were also determined by carrying out the summation up to $l_{max}^{M}=5$. The magnetic force theorem was applied as described in Ref. [9] to calculate inter-atomic exchange interactions as formulated by Liechtenstein [10] and using an extended set of k-points for the Brillouin zone integration. We also calculated the electronic structure and the exchange interaction in the paramagnetic state by modelling it within the Disordered Local Moment (DLM) approximation according to Györffy *et al.* [11].

## 3. Results and Discussion.

The FPLAPW calculation has suggested spin-polarized magnetic ground state of MnB$_4$ with magnetic moment of 0.53 $\mu_B$/f.u and negligibly small moment within Boron muffin-tin spheres. Thus it is fully associated with moment of Mn. In the figure 2 we plot the calculated spin-polarized Density of States (DOS).

[figure 2 here]

The magnetic instability is associated with a sharp (spin-split in the figure) peak near the Fermi level due to Mn d-metallic states hybridized along Mn chains. To investigate the character of the MnB$_4$ magnetism further we performed KKR-ASA calculations to employ DLM formalism for modeling the paramagnetic state, and in order to estimate from first-principles the exchange coupling between Mn moments within the framework of Green Function based magnetic force theorem [10]. The KKR-ASA spin polarized calculations has lead to the similar results as FPLAPW – calculated moment of ferromagnetic state is 0.36 µ$_B$/Mn. The observed disagreement is usual for open crystal systems and related to the covalent character of Mn-B bonds, wchich are not well treated treated by ASA approximation (see discussion on this point for the case of relative MnB$_2$ material in Ref.[2]). The DLM calculation, however, leads to the zero Mn moment and thus suggests a very itinerant character of the Mn magnetism on contrast for example to the MnB$_2$, where the moments are relatively rigid. The calculated exchange constants $J_{ij}$ for Heisenberg Hamiltonian $H = \sum_{ij} J_{ij} \vec{e}_i \vec{e}_j$, where $\vec{e}_i$ is unit vectors of spin at the *i*-th Mn site reveals essentially quasi one-dimensional character of magnetism. The only significant interaction is ferromagnetic between nearest-neighbour Mn moments along c-axis - *J(1NN) = 0.29 mRy*. All other interactions constants are almost two order of magnitude smaller and their values is on the verge of computational accuracy of our scheme (0.01 mRy).

In the figure 3 we plot calculated (FLAPW) band structure along selected directions in the reciprocal space. Only one metallic band crosses the Fermi level. Four Mn valent d-electrons are forms p-d hybrid bands below Fermi level. Thus it appears that four valence Mn d-electrons participate in stabilization of the boron network, whereas remaining one participates in formation of one-dimensional metallic bands along chains of Mn atoms (see figure 1). These metallic chains are electronically separated from each other by the boron network. It suggests that MnB$_4$ may be considered as a candidate for being a model system to study 1D on-.band Hubbard model.

## 4. Conclusions.

The results of first-principles calculations presented here suggest that MnB$_4$ has magnetic ground state (with moment at about 0.5μ$_B$/Mn) such that magnetic interactions along the Mn-chains are almost two orders of magnitude larger that inter-chain interactions. Moreover, we have find an indication of 1D character of the metallic state in this system. The rather small value of the magnetic moment and its vanishing in the Disordered Local Moment state suggest strong itinerancy of the moments. All together make this system a good candidate for further experimental studies in low temperature regime.

**Fig. 1.** Crystal structure of MnB$_4$. 1D Mn chaines are perpendicular to the plane of the figure. Small balls are Boron atroms. Large balls inside the Boron cages is Mn

**Fig.2.** Calcualted spin polarized Density of State of MnB$_2$ (FLAPW). Total and Mn atom projectes.

**Fig. 3**. Calculated band structure along selected directions in reciprocal space. Upper panel spin-up band. Lower panel spin down band.

**Fig. 1.**

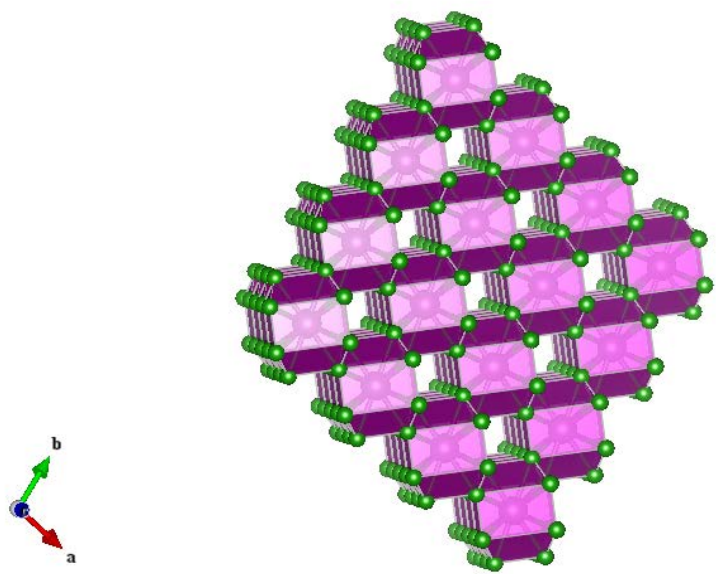

**Fig. 2.**

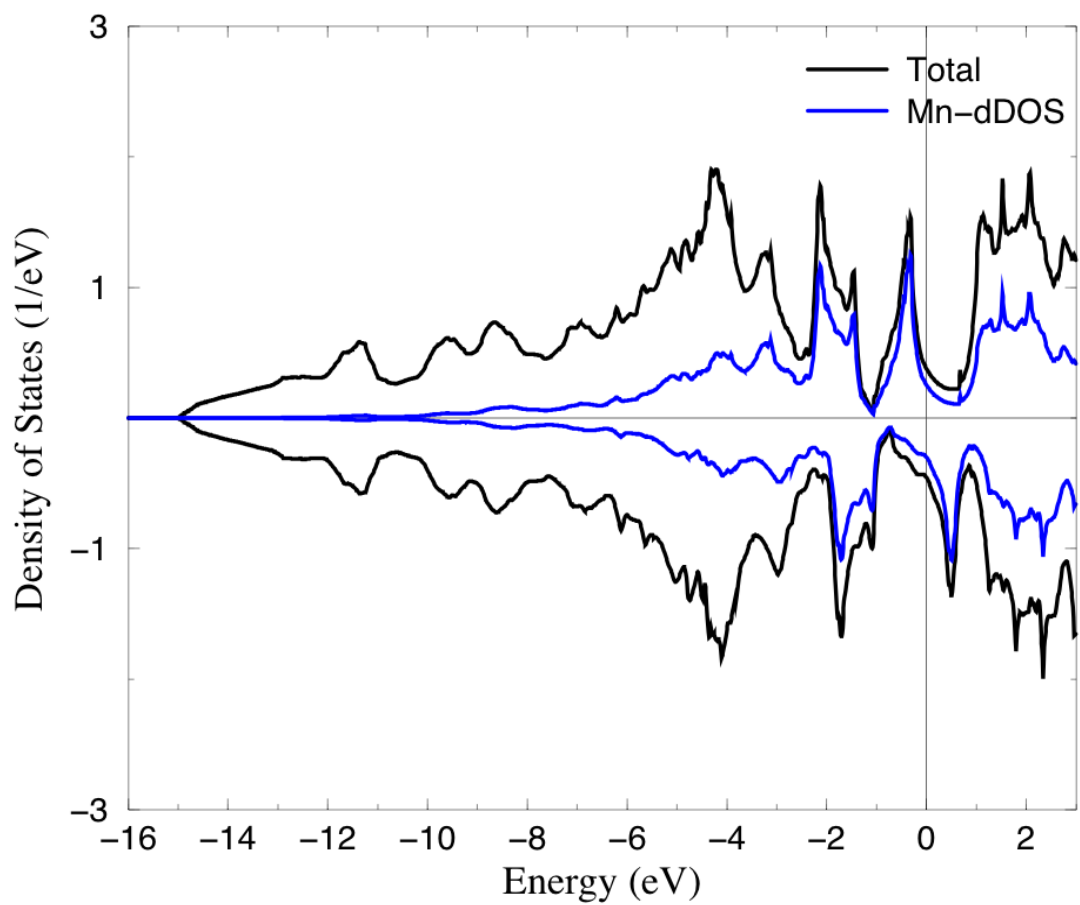

**Fig. 3.**

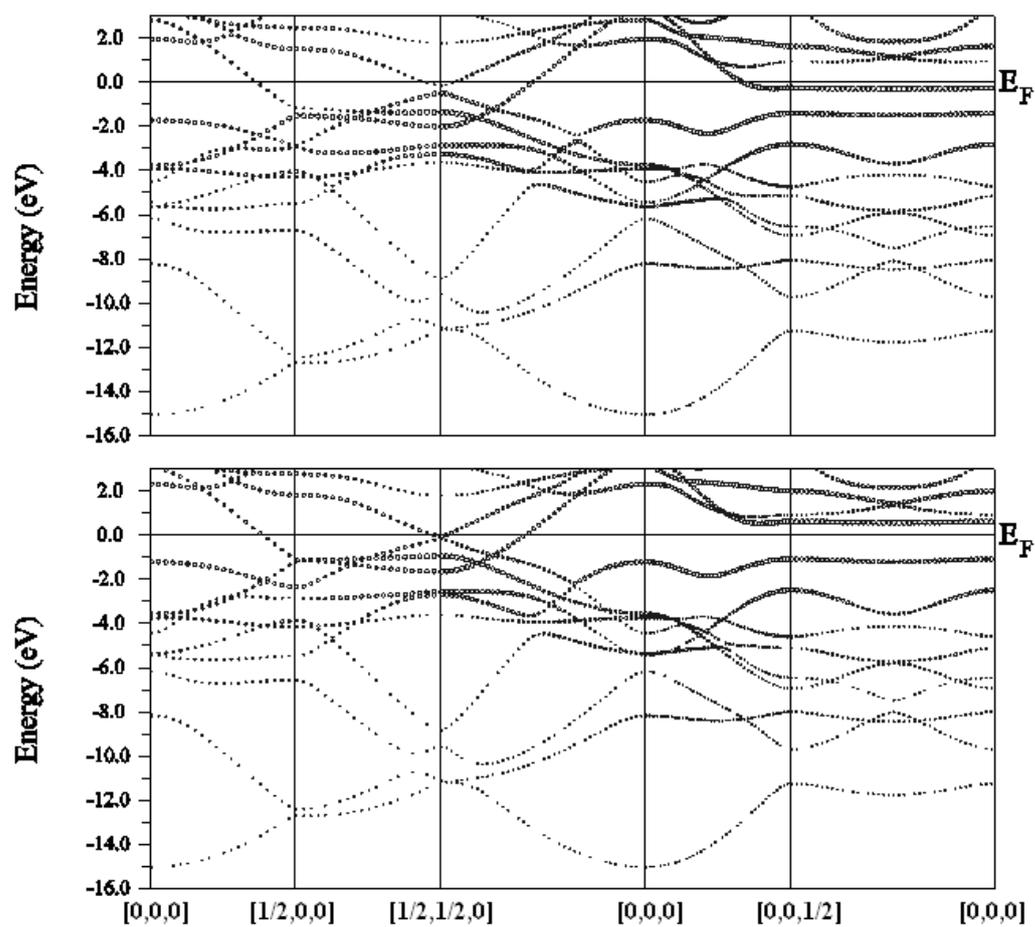